\documentclass[twoside]{article}
\usepackage{graphicx}
\usepackage{fancyhdr}
\usepackage{multicol}
\usepackage{styl-ap}

\begin{document}
\title{SW\,Sex stars, old novae, and the evolution of cataclysmic variables}
\author{Linda Schmidtobreick\work{1}, Claus Tappert\work{2}}
\workplace{
European Southern Observatory, Casilla 19001, Santiago 19, Chile
\next
Departamento de F\'\i sica y Astronom\'\i a, 
Universidad de Valpara\'\i so, Avda.\ Gran Breta\~na 1112, 
Valpara\'\i so, Chile
}
\mainauthor{lschmidt@eso.org}
\maketitle

\begin{abstract}%
The population of cataclysmic variables with orbital periods right above the
period gap are dominated by systems with extremely high mass transfer rates,
the so-called SW Sextantis stars.
On the other hand, 
some old novae in this period range which are
expected to show high mass transfer rate instead 
show photometric and/or 
spectroscopic resemblance to low mass transfer systems like dwarf novae.
We discuss them as candidates for so-called hibernating systems, CVs that 
changed their mass transfer behaviour due
to a previously experienced nova outburst. This paper is designed to
provide input for further research and discussion as the results as such are 
still very preliminary.

\end{abstract}


\begin{multicols}{2}
\section{Introduction}
In the following we briefly review our current knowledge on the populations
of old novae and SW Sextantis stars especially in the context of the evolution
of cataclysmic variable stars (CVs).
\subsection{Evolution of CVs}
In general, the evolution of stable interactive binaries is driven by 
angular
momentum loss which as such determines the mass transfer rate.
The loss of angular momentum as a source for stable mass transfer implicates
that the general evolutionary direction for CVs is from longer to
shorter orbital periods. According to the 
standard model of CV evolution, the main source of angular momentum loss
for CVs with orbital periods $P_{\rm orb} > 3$\,h is magnetic braking.
Due to the continuous mass transfer, the secondary is pushed out of 
thermal equilibrium and becomes bloated.
At a period of about 3\,h, i.e. at the upper edge of the period gap,
the secondary becomes fully convective, magnetic 
braking ceases and only the much weaker braking through the emission of 
gravitational radiation continues.
Therefore, the mass transfer is greatly diminished which allows the secondary star
to relax into a state of thermal equilibrium and to contract to a 
volume corresponding to its mass.
It thus loses contact with its Roche lobe and mass transfer stops completely. 
The angular momentum loss via gravitational radiation shrinks the orbit of
the now detached and therefore hardly detectable binary 
until the secondary star fills its Roche lobe 
again at an orbital period of about 2\,h, i.e. at the lower edge of the gap
and the binary continues as a low mass transfer CV below the gap.
For an extensive review of the current 
understanding of CV evolution, see Knigge et al.\ (2011).

\subsection{SW\,Sex stars}
This sub-class was originally defined by Thorstensen et al.\ (1991). 
incorporating eclipsing nova-like stars with single-peaked emission lines, high velocity line 
wings, strong He\,II emission but no polarisation, and transient absorption
features at orbital phases around $\phi = 0.5$. The radial velocity curves
show an offset of 0.2 cycle with respect to the phase defined by the eclipse.
Today, SW\,Sex stars are considered novalike stars with an extremely high
mass transfer rate  (see e.g. Rodr\'\i guez-Gil et al.\, 2007a,b). This is supported
by the high temperatures found for the white dwarfs in these 
systems (Townsley \& G\"ansicke, 2009). 
While SW\,Sex stars  used to be considered as rare and strange objects with
unusual behaviour, it has been shown in the last years that they 
in fact represent the 
dominant CV population in the orbital period range between about three 
and four hours (Schmidtobreick et al., 2012). From a sample of CVs that was purely seleceted to have an orbital period between three
and four hours, they find that the percentage of
SW\,Sex stars in this range must exceed 85\%. 
According to the standard model, all long period CVs 
have to evolve through this period range before entering the gap
and will thus share the SW\,Sex characteristics during that time.
This makes the SW\,Sex phenomenon a phase in the secular evolution of CVs
(Schmidtobreick et al.\ in preparation). 

\subsection{Old Novae}
Classical novae are CVs that experience a thermonuclear runaway on the 
surface of the white dwarf where the accreted material has reached a 
critical mass. In this process, the binary is not destroyed and mass transfer 
is usually re-established within a few months. 
It is therefore safe to assume that
nova explosions are recurrent events and part of the evolution of every CV
with a sufficiently high mass transfer rate to accumulate the necessary material
on the white dwarf surface within its lifetime. 
Inbetween the nova events, the behaviour of the
CV depends on properties like orbital period, mass-transfer rate,
and magnetic field of the white dwarf that also determine its subtype.
In addition, the hibernation model predicts changes of the mass transfer
rate in the evolution of the pre- and post-nova: 
after an initial phase of high mass transfer rate -- due to the secondary 
star being driven strongly out of thermal equilibrium via irradiation from
the eruption-heated white dwarf -- which increases the 
distance between the two stars, the binary should descend into a long
state of low mass transfer once the white dwarf cooled down and the secondary
is allowed to relax more towards thermal equilibrium. 
(Shara et al., 1986; Prialnik \& Shara, 1986).
Some potential evidence for hibernation has been presented 
in the form of old nova shells around CVs that have
previously been known as low mass transfer systems, 
i.e. Z\,Cam (Shara et al., 2007)
and AT\,Cnc (Shara et al., 2012). 
However this interpretation is not exclusive,
as the presence of nova shells
around dwarf novae could also indicate 
that all types of CV (including dwarf novae) can experience 
a nova explosion during their lifetime without necessarily undergoing cyclic
changes of the mass transfer rate.

\section{What can we learn combining our knowledge from old novae and SW\,Sex stars?}
Several years ago, we conducted a project investigating
old novae which had experienced large outburst
amplitudes (Schmidtobreick et al., 2003; 2005).
The idea behind this was that since the
absolute magnitude of a nova explosion depends mainly on the mass of the
white dwarf (Livio, 1992) it thus differs only slightly for different
systems. Thus, novae with large outburst amplitudes are likely to be
intrinsically faint CVs and therefore candidates for low mass transfer rate
systems. 


Two systems of our sample, V842 Cen and XX Tau, show spectroscopic properties
that, compared to other old novae, indicate a rather low mass transfer rate. 
The spectra present comparatively strong Balmer emission lines from H$\alpha$
bluewards down to H11. A weak HeII emission line is present at 469 nm and
indicates a hot component in the system, but at the same time the presence
and strength of the HeI series is evidence for a significant amount of cooler 
material. We note that most old novae do not share these characteristics
(e.g., Ringwald et al. 1996, Tappert et al. 2012, 2014). For XX Tau, 
preliminary analysis of long-term photometric monitoring covering a range of
150\,d furthermore shows a probable dwarf-nova like outburst with an
amplitude of $\sim$0.8 mag and a duration of $\sim$10 d (Tappert, private
communication), which represents additional evidence for a comparatively low
mass transfer rate.

%
The approach by Tappert et al.\ (2012) to re-discover lost old novae revealed
two more such candidates from spectroscopy: V2109\,Oph and V728\,Sco. 
In particular V728\,Sco
which has also been observed photometrically can be considered a 
low mass
transfer system as it also shows frequent stunted dwarf novae outbursts
(Tappert et al., 2013a). 

For three of these low mass transfer old novae, i.e. V842\,Cen, V728\,Sco, and XX\,Tau, at least
rough values for the orbital period could be established 
(Woudt et al., 2009; Luna et al., 2012; Tappert et al., 2013a; 
Rodr\'\i guez-Gil \& Torres, 2005)
and they all fall into the range
of SW\,Sex stars, i.e. between 3 and 4 hours. 

On the one hand, this period range is thus dominated
by high mass transfer CVs. On the other hand, also the majority of old novae 
have high mass transfer rates. In fact, the period distribution of old novae
shows a distinctive peak in the 2.8-5 h period range, while few novae are
found in the period bins that are dominated by dwarf novae (e.g., Tappert et 
al. 2013b, Schmidtobreick \& Tappert 2014). Thus, the indicated low mass 
transfer nature of V842 Cen, V728 Sco and XX Tau is somewhat surprising. 


For the sake of completeness, we point out 
that other low mass transfer novae are known. For example, 
the system V446 Her (Porb = 4.97 h, Thorstensen \& Talor, 2000) 
shows frequent "stunted" dwarf-nova like outbursts (Honeycutt et al. 1998). 
However, we here restrict our discussion to systems with orbital periods 
in the 3-4\,h range.

When thinking about a reason why several old novae would appear 
as low mass 
transfer systems even though some of them are even situated in the 
SW\,Sex regime, hibernation comes naturally to mind.
It seems an attractive explanation that these CVs have indeed been
SW\,Sex stars which due to the nova explosion in the recent past 
were pushed into a
low state and thus experience low mass transfer rates at the moment.
In this context we point out that V728\,Sco in outburst presents a triangular 
eclipse shape that is typical for SW Sex stars (Tappert et al., this volume).

\section{Conclusions and future work}

CVs with orbital periods between 3 and 4 hours are in general
SW\,Sex stars and as such experience high mass-transfer rates. 
However, we find
evidence for three old novae with orbital periods in this range that
should be of SW\,Sex type but are instead low mass transfer systems.
We tentatively conclude that the mass transfer of these systems was changed by the 
nova eruption, similar to what is proposed in the hibernation scenario.
Whether the mass transfer in these systems will completely cease or 
just remain on a low level before rising up again
remains to be investigated.
To better understand the recent mass transfer history and to thus shed light
on the evolutionary state of these systems, it would be important to measure
the dynamical masses and the temperatures of the two components in
order to determine the binary solution. For this, it is valuable to have
a system like V728\,Sco in the sample which as eclipsing binary offers
more opportunities to determine the binary parameters. 

To test the idea that the low mass transfer novae in the
SW\,Sex regime are affected by hibernation, it will be essential to check
for the few existing dwarf novae in the SW\,Sex range whether they 
have experienced a nova eruption in the past. We thus started an observational
project to look for nova shells around these objects. 
Other possible candidates for hibernating CV could be found among
the so-called pre-CVs, white-dwarf/main-sequence binary which are 
detached systems. Examples for such candidates are 
LTT\,560  (Tappert et al., 2011) and QS\,Vir 
(Drake et al., 2014, and references therein)
which both have orbital periods in
the SW\,Sex range and show evidence of accretion.
If indeed a significant number of these 
low mass transfer systems in the SW\,Sex regime can be shown to 
have experienced a nova outburst in the
past, this would not only be strong evidence for the hibernation 
scenario but at the same time also prove the
evolutionary significance of the SW\,Sex stars as it would yield 
a natural explanation for the few remaining CVs
in the 3-4\,h period range that do not follow the SW Sex behaviour.

\thanks
This research was supported by FONDECYT Regular grant
1120338 (CT). We gratefully acknowledge the use of the SIMBAD database, operated
at CDS, Strasbourg, France, and of NASA's Astrophysics Data System Bibliographic
Services. We thank an anonymous referee for valuable comments.

\bigskip
\bigskip
\noindent {\bf DISCUSSION}

\bigskip
\noindent {\bf VITALY NEUSTROEV:} You claim that some old novae have a very
small mass transfer rate. Does it mean that these stars should show 
DN-type outbursts?

\bigskip
\noindent {\bf LINDA SCHMIDTOBREICK:} Yes, indeed. And some stars do show 
such outbursts. We don't yet have a lot of data but we have observed
several outbursts for V728\,Sco. 

\end{multicols}
\end{document}